\documentclass[reprint,
 amsmath,amssymb,
 aps,
 pra,
]{revtex4-2}

\usepackage{graphicx}
\usepackage{dcolumn}
\usepackage{bm}
\usepackage{braket}
\usepackage{color}

\begin{document}

\preprint{APS/123-QED}

\title{Compression Causes Expansion and Compaction of Jammed Polydisperse Disks}

\author{Daisuke S. Shimamoto}\email{daisuke\_shimamoto@seas.harvard.edu}
\affiliation{
Center for Interdisciplinary Theoretical and Mathematical Sciences (iTHEMS),\\
 RIKEN, Wako 351-0198, Japan
}
\author{Miho Yanagisawa}
 \email{myanagisawa@g.ecc.u-tokyo.ac.jp}
 \altaffiliation[Also at ]{Center for Complex Systems Biology, Universal Biology Institute, The University of Tokyo., Graduate School of Science, The University of Tokyo, Hongo 7-3-1, Bunkyo, Tokyo 113-0033, Japan}
\affiliation{
Komaba Institute for Science, Graduate School of Arts and Sciences, The University of Tokyo, Komaba 3-8-1, Meguro, Tokyo 153-8902, Japan
}

\date{\today}

\begin{abstract}
The compression of materials is generally expected to increase their density through compaction. In contrast, recent studies have shown that in certain structured materials, transitions between metastable states can lead to counterintuitive expansion. Particle assemblies also experience transitions between metastable states during the cycles of compression and decompression, and they are generally assumed to cause compaction. Here, we report the expansion of highly polydisperse soft disks in two dimensions following a cycle of compression and decompression. We compare the jamming densities before and after cycles with different maximum area fractions, $\phi_{\rm max}$. While cycles with small amplitude cause compaction as expected, cycles with large amplitude decrease the jamming density and lead to expansion. We demonstrate that this expansion originates from size segregation, which reduces the volumetric modulus and changes the local size distribution. These findings reveal an expansion in a disordered particle assembly and demonstrate that polydispersity can alter the outcome of rearrangements between metastable states.

\end{abstract}

\maketitle

\section{Introduction}
Materials are generally expected to shrink after compression beyond the elastic range~\cite{batey2009soil,nawaz2013soil,chaudhuri2010jamming,vaagberg2011glassiness,guiochon1995consolidation,cooper1962compaction,vu2019soft}, but recent studies have shown that structured materials~\cite{zha2024reversible,nicolaou2012mechanical,ducarme2025exotic,nicolaou2024metamaterials,qu2017experiments} can violate this intuition through responses called negative compressibility or counter-snapping. In such systems, compressive stress can induce expansion~\cite{zha2024reversible,nicolaou2024metamaterials,qu2017experiments}, or tensile stress can cause compaction~\cite{ducarme2025exotic,nicolaou2012mechanical}. These responses often emerge from transitions between metastable states, even when the interactions or elasticity themselves are not exotic. This motivates us to ask whether a cycle of compression and decompression can make a jamming of particles less dense.

Compaction induced by compression is observed in a wide range of systems, including soils~\cite{batey2009soil,nawaz2013soil}, colloids~\cite{guiochon1995consolidation}, powders~\cite{cooper1962compaction}, and foams~\cite{vu2019soft}. In granular materials, compression commonly causes compaction. Even in athermal frictionless models, the jamming density depends on the preparation and loading history. Isotropic deformation proceeds through irreversible non-affine rearrangements and stress drops, and the cycles of compression and decompression generally drive the system toward more densely jammed states~\cite{kumar2016memory,matsuyama2021geometrical,luding2021does,luding2023elastic}. The reported responses are consistent with compaction. However, expansion induced by compression has rarely been considered in this context. Following the paper~\cite{matsuyama2021geometrical}, we hereafter use the term mechanical annealing for such an athermal quasistatic compression-decompression cycle, which drives transitions between mechanically stable states.

Reduction in density and training toward a different jammed state are nevertheless known under other protocols, such as shear~\cite{kumar2016memory,babu2021dilatancy}, and tapping or vibration through repeated rearrangements under gravity~\cite{knight1995density,philippe2002compaction,pugnaloni2008nonmonotonic}. These responses show that complex protocols can either increase or decrease density.

Strong size polydispersity can give rise to phenomena distinct from those observed in systems with small polydispersity. The density and elastic response depend on whether small particles merely occupy voids as rattlers or participate in the contact network~\cite{prasad2017subjamming,petit2020additional,kumar2016tuning,hara2021phase}. Further, the jamming density of polydisperse particles depends on the span and shape of the size distribution because these parameters control coordination and the force network~\cite{voivret2007space,estrada2016effects,shimamoto2023common,meer2024estimating,saitoh2025jamming}. Mechanical annealing of highly polydisperse systems may therefore produce a qualitatively different response from compaction.

In this study, we investigate how an athermal quasistatic compression-decompression cycle changes the jamming density of highly polydisperse soft disks with a power-law size distribution. We compare the jamming densities before and after cycles over a wide range of $\phi_{\rm max}$. While cycles with small amplitude lead to compaction as expected, cycles with large amplitude decrease the jamming density and lead to expansion. We show that this expansion is accompanied by size segregation, which locally reduces particle-size polydispersity, lowers the volumetric modulus, and facilitates the formation of contacts between similarly sized particles. In highly polydisperse systems, cycles of compression and decompression with large amplitude can induce size segregation, which can overcome the conventional compaction and can lead to expansion.
This is a response enabled by the size polydispersity and cycles with large amplitude studied here. In particular, our bidisperse control has a narrow radius ratio of $1:1.4$, so the absence of expansion in that system should not be generalized to the bidisperse mixtures with larger size ratios. These results reveal that loading-induced mechanical annealing can expand highly polydisperse soft-particle assemblies rather than compact them.

\begin{figure}
    \centering
    \includegraphics[width=8.6cm, bb = 0 0 451 140]{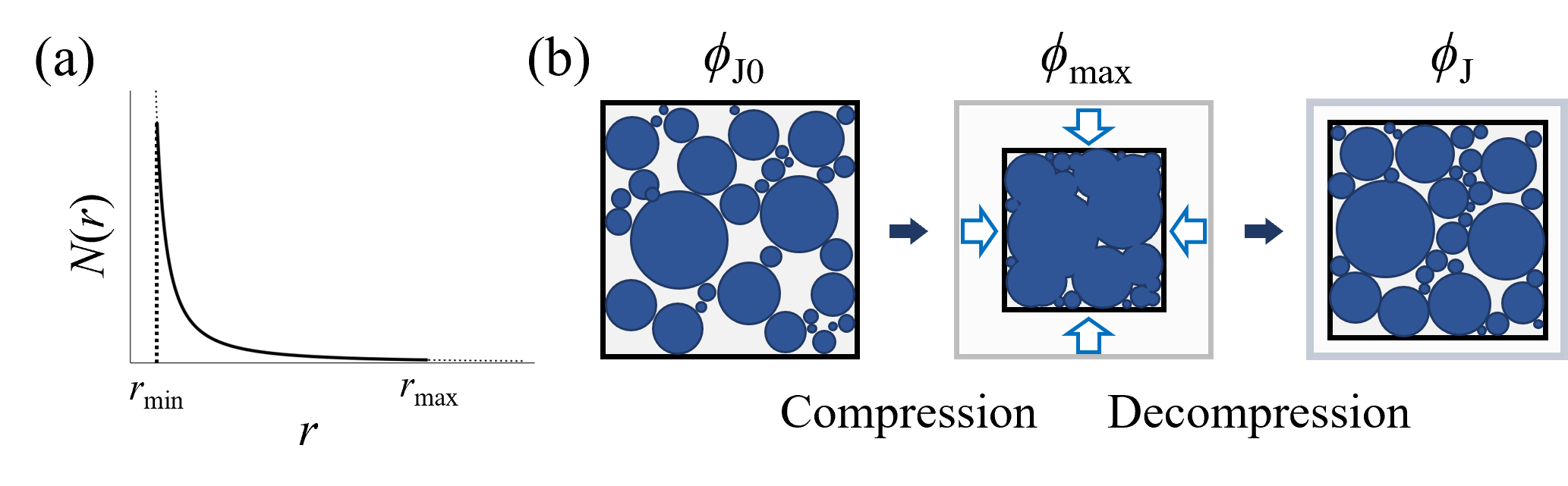}
    \caption{Schematic showing (a) the power-law size distribution of disks and (b) the athermal quasistatic compression-decompression cycle. Randomly placed disks below the jamming point are compressed isotropically and then decompressed immediately when the density reaches $\phi_{\rm max}$. The density immediately after the pressure increases to a finite value during compression is defined as $\phi_{\rm J0}$, and that immediately before the pressure reaches zero during decompression is defined as $\phi_{\rm J}$.}
    \label{cycle}
\end{figure}

\section{Model}

We consider a two-dimensional jammed polydisperse soft disks. The disks interact through purely repulsive forces only when they overlap. The interaction potential between disks $i$ and $j$ is the harmonic potential
\begin{equation}
U=\epsilon\left(r_{ij}-d_{ij}\right)^{2}\Theta\left(r_{ij}-d_{ij}\right),
\end{equation}
where $d_{ij}$ is the distance between the centers of the $i$-th and $j$-th particles, $r_{ij}$ is the sum of their radii, and $\epsilon$ has the dimension of energy per length squared. Since the system size $L$ sets the length scale, $\epsilon L^2$ defines the unit of energy in our simulations.
Further, $\Theta(x)$ is the Heaviside step function, defined as $\Theta\left(x\right)=1$ for $x>0$, and $\Theta\left(x\right)=0$ otherwise. The unit of length was set to be the system size. The pressure is calculated as
\begin{equation}
p=\frac{1}{2L^2}\sum_{i<j}d_{ij}f_{ij}.
\end{equation}
For the harmonic interaction, the overlap $\delta_{ij}=r_i+r_j-d_{ij}$ gives the normal contact force $f_{ij}=2\epsilon\delta_{ij}$. The dimensionless pressure is normalized as $P=p/\epsilon$, which is used in the plots.

The distribution of particle radii, $N(r)$, was set to either a power distribution with cutoffs (Fig.~\ref{cycle}(a)) or a bidisperse distribution. The power size distribution is expressed as:
\begin{equation}
N(r)=
\frac{a-1}{r_{\rm min}^{-a+1}-r_{\rm max}^{-a+1}}r^{-a}
\Theta\left(r-r_{\rm min}\right)
\Theta\left(r_{\rm max}-r\right),
\end{equation}
where $r_{\rm min}$ and $r_{\rm max}$ are the lower and upper cutoffs of the distribution, respectively, and $a$ is the exponent that characterizes the shape of the distribution.
Particles following these distributions were generated using the inverse function method~\cite{devroye2006nonuniform} in the same way as we have previously reported~\cite{shimamoto2023common}.
At the initial jamming point, the largest particle radius was at most 0.15$L$, and it was at most 0.33$L$ even at $\phi=5$, so that the box length always exceeded the diameter of the largest particle.
The bidisperse distribution is an equimolar binary mixture of particles with a size ratio of $1$:$1.4$. The total number of particles was typically set to $4000$. For $a=2.7$, the number of particles was set to $16000$ to ensure a size range spanning two orders of magnitude by generating a sufficient number of the minor component, large particles.
The \textit{density}, $\phi$ was defined as the sum of the areas of all particles divided by the total area of the simulation cell:
\begin{equation}
\phi = \frac{1}{L^2}\sum_{i=1}^{N} \pi r_i^2,
\end{equation}
where $L$ is the side length of the square cell and $r_i$ is the radius of the $i$-th particle.

Initial states were prepared by relaxing the random configuration generated from the uniform distribution. The particle coordinates were drawn independently from a uniform distribution in a square cell with periodic boundary conditions. Thus, no size segregation was imposed in the initial configurations. The system was then relaxed to mechanical equilibrium using the FIRE algorithm~\cite{bitzek2006structural} until the maximum net force on any particle was below $10^{-13}$. The initial density was set to 0.91 for $a=2.306$ and 2.7, 0.88 for $a=1.8$, and 0.835 for bidisperse, below an estimated jamming density obtained from preliminary simulations.

We applied a quasistatic cycle of compression and decompression by uniformly increasing and subsequently decreasing the particle radii while keeping $L$ fixed. The initial state was isotropically compressed in increments of $\delta\phi=10^{-4}$ until the density reached $\phi_{\rm max}$. After every compression and decompression step, the system was relaxed to mechanical equilibrium using FIRE. Thus, the protocol introduces neither thermal fluctuations nor a finite temperature. Smaller increments did not significantly affect the results. We define $\phi_{\rm J0}$ as the density at which the pressure first becomes finite during compression. Once $\phi$ reached $\phi_{\rm max}$, decompression began in increments of $\delta\phi=-10^{-4}$, and $\phi_{\rm J}$ was defined as the density immediately before the pressure dropped to zero. We then compared $\phi_{\rm J0}$ and $\phi_{\rm J}$ (Fig.~\ref{cycle}(b)). Larger $\phi_{\rm max}$ corresponds to a larger-amplitude cycle. To reduce statistical uncertainty, all results were averaged over 10 independent trials.
\begin{figure}
    \centering
    \includegraphics[width=8.6cm, bb = 0 0 2200 1300]{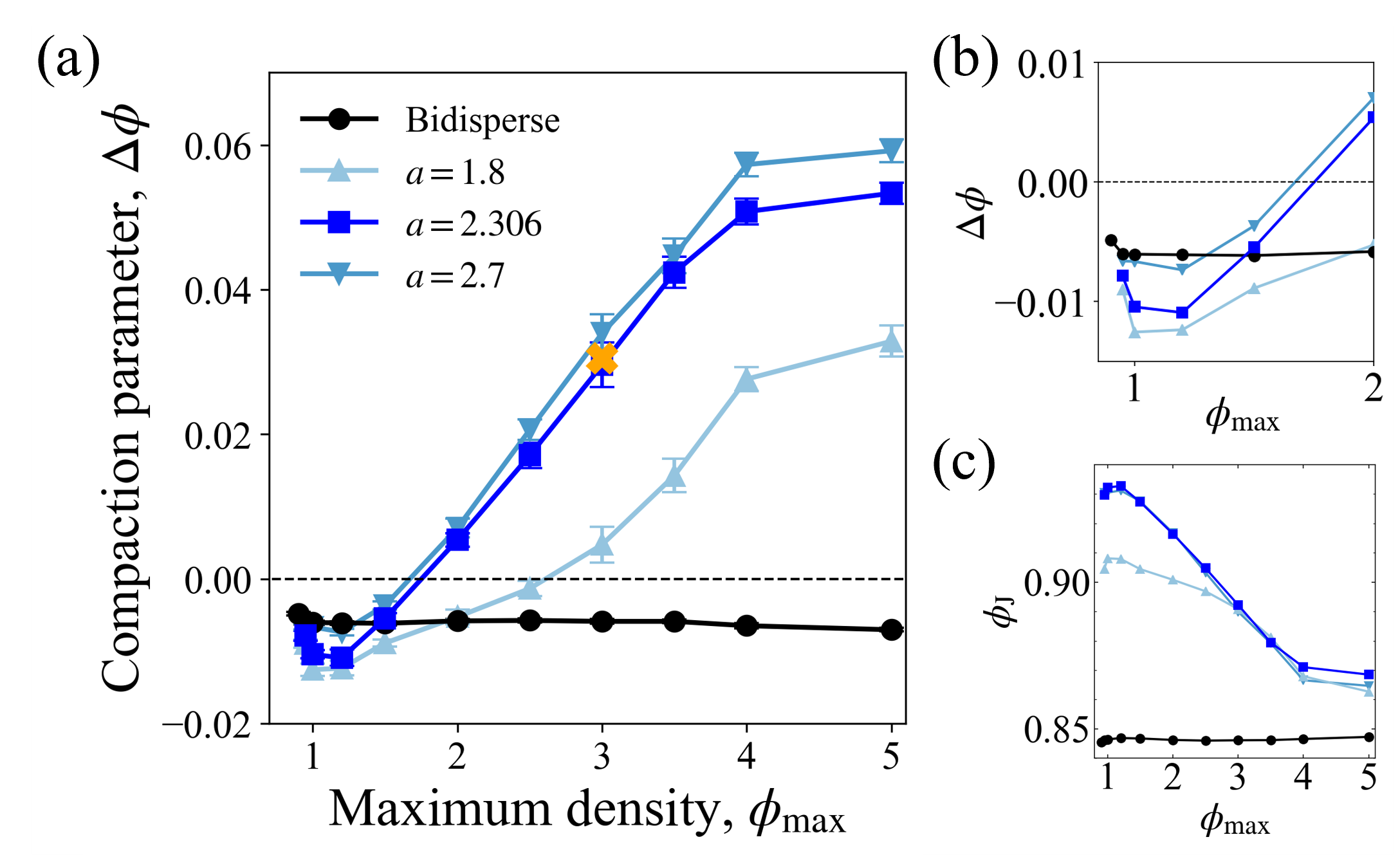}
    \caption{(a) Plot of $\Delta\phi_{\rm}$ against $\phi_{\rm max}$, and (b) its enlarged plot for $\phi_{\rm max}<2$. The result without FIRE is also shown in (a) with an orange crossing. (c) Plot of $\phi_{\rm J}$ against $\phi_{\rm max}$. Each marker corresponds to results from different parameter settings and is not a time-series plot of the same trial. The results for Bidisperse systems are indicated as circles, and polydisperse systems are as upward triangles, squares, and downward triangles for exponents $a = 1.8$, $2.306$, and $2.7$, respectively. The error bars indicate the standard error.}
    \label{inflation}
\end{figure}
The volumetric modulus near the jamming point, $\kappa=dP/d\phi$, was obtained from the pressure histories as the slope of $P(\phi)$ for $0<P\le0.05$.

\section{Results}

\subsection{From compaction to expansion}

Figure~\ref{inflation} shows the change in density across the compression-decompression cycle, $\Delta\phi=\phi_{\rm J0}-\phi_{\rm J}$, for the bidisperse and polydisperse systems ($a=1.8, 2.306, 2.7$), plotted against the maximum density, $\phi_{\rm max}$.
The absolute value of $\phi_{\rm J0}$ depends systematically on the particle-size distribution because the degree of polydispersity affects the efficiency with which smaller particles fill the voids between larger particles. Thus, a fixed $\phi_{\rm max}$ corresponds to slightly different overcompression amplitudes, $\phi_{\rm max}-\phi_{\rm J0}$, for the different distributions. Nevertheless, because $\Delta\phi=\phi_{\rm J0}-\phi_{\rm J}$ is calculated separately for each realization, it measures the cycle-induced change relative to the initial jamming density of that realization rather than a difference between the absolute jamming densities of distinct particle-size distributions. We therefore compare the qualitative response within each distribution and do not assume quantitative equivalence among distributions at the same absolute $\phi_{\rm max}$.
A negative $\Delta\phi$ indicates compaction, whereas a positive value indicates expansion. We first describe the results for the bidisperse system as a reference. In the bidisperse system, $\Delta\phi$ was consistently negative, regardless of $\phi_{\rm max}$. In the region where $\phi_{\rm max}$ was small ($<1$), $\Delta\phi$ gradually decreased and reached a plateau for $\phi_{\rm max}$ of unity. This compaction behavior, reported by Kumar et al.~\cite{kumar2016memory}, is an example of the history dependence of the jamming point. 

Next, we present the results for the polydisperse systems, which are the main focus of this study. In contrast to the narrow bidisperse system considered here, the highly polydisperse systems exhibited expansion after large-amplitude cycles. As an example, we describe the results for $a=2.306$ (squares in Fig.~\ref{inflation}). For small $\phi_{\rm max}$ ($\phi_{\rm max}<1$), $\Delta\phi$ decreased as $\phi_{\rm max}$ increased, indicating compaction. However, once $\phi_{\rm max}$ exceeded unity, $\Delta\phi$ began to increase instead of leveling off. When $\phi_{\rm max}>1.5$, $\Delta\phi$ became positive and continued to increase until it plateaued at approximately $\Delta\phi=0.04$. Thus, increasing the cycle amplitude produced a transition from compaction to expansion, which we term negative compaction. The final area fraction ranged from approximately $93\%$ for $\phi_{\rm max}=1.2$ to $86\%$ for $\phi_{\rm max}=5.0$. The same change from negative to positive $\Delta\phi$ was observed for $a=1.8$ and $a=2.7$ (upward and downward triangles in Fig.~\ref{inflation}).
To evaluate the impact of using the FIRE algorithm on the results, we performed control simulations at $\phi_{\rm max} = 3.0$ by numerically integrating the overdamped equations of motion without employing it (a crossing in Fig.~\ref{inflation}(a)). We confirmed that the results averaged over five independent runs were consistent with those obtained using FIRE, with discrepancies in $\Delta\phi$ less than 4\%.

\subsection{Size segregation}

\begin{figure}
    \centering
    \includegraphics[width=8.6cm, bb = 0 0 334 466]{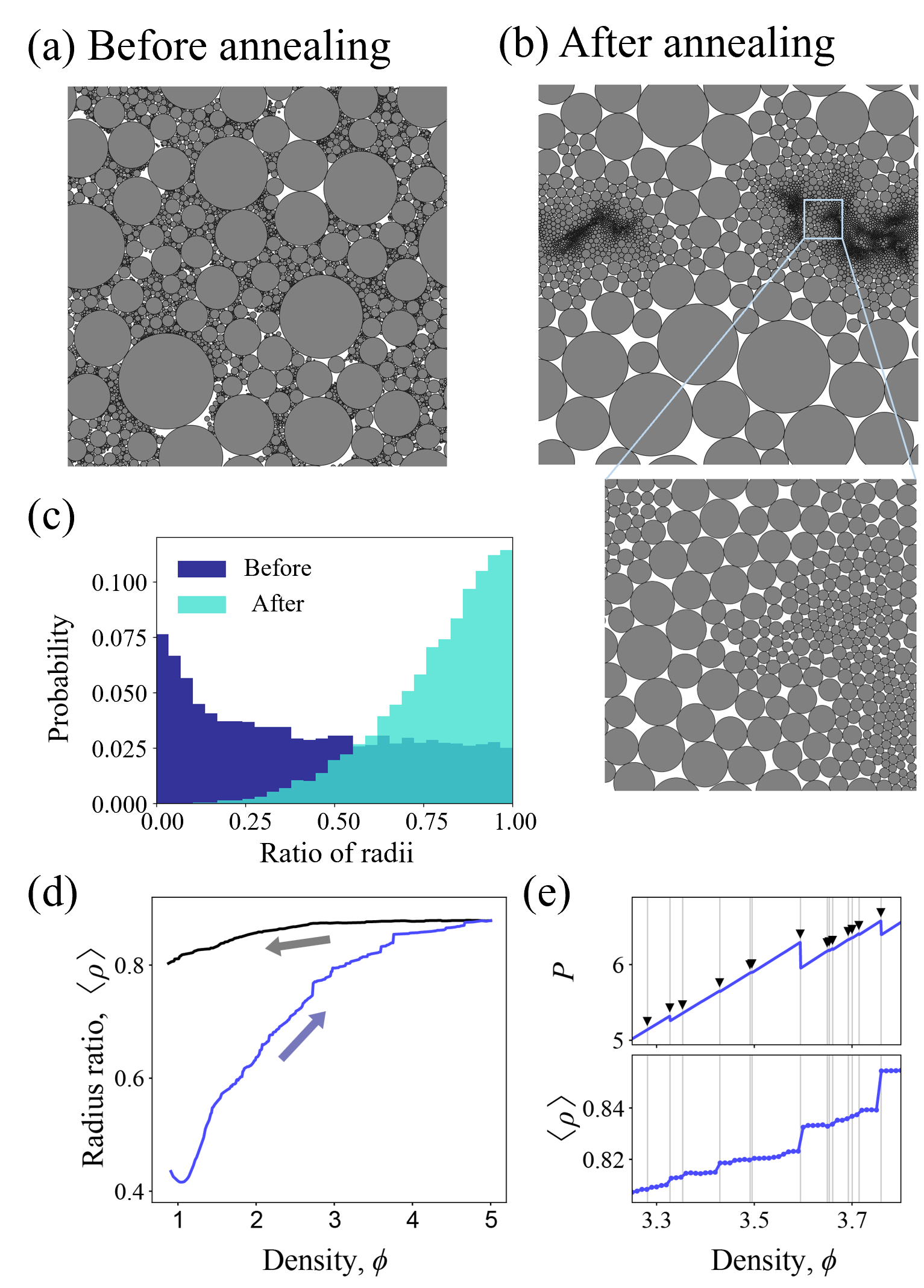}
    \caption{(a,b) Examples of packing of polydisperse particles ($a=2.306$) at the jamming point before (a) and after (b) the cycle of compression
    with $\phi_{\rm max}=5$. The area enclosed by the blue box in (b) is magnified. The darker appearance in (b) is owing to the outlines of accumulated small particles and does not indicate a high local density. (c) Probability distribution of radius ratios between the particles in the nearest neighbors. The radius ratios are calculated as the smaller radius divided by the larger one.
    (d) Mean radius ratio between neighboring particles in terms of Laguerre-Voronoi tessellation as a function of density during compression (blue) and decompression (black) for $\phi_{\max}=5$. The arrows indicate the direction of the process. (e) Larger view of a representative interval during compression in (d). The upper and lower panels show the pressure and mean radius ratio, respectively. Triangles indicate pressure drops, and the vertical gray lines indicate their positions.}
    \label{segregation}
\end{figure}

To elucidate the cause of this negative compaction, we examined changes in the packing configuration at the jamming point before and after the compression-decompression cycle. Figure~\ref{segregation} shows the probability distribution of the radius ratio between nearest-neighbor particles, obtained from the packings before (Fig.~\ref{segregation}(a)) and after (Fig.~\ref{segregation}(b)) the cycle.

To quantify the structural changes induced by the cycle, we analyzed the distribution of size ratios among nearest-neighbor particles (Fig.~\ref{segregation}(c)). Nearest neighbors were identified using the Laguerre-Voronoi tessellation, which accounts for particle size when defining local neighborhoods~\cite{matousek2013lectures,Hart2020swX}. The size ratio was defined as the smaller radius divided by the larger one for each neighbor pair. Before the cycle, the distribution was broad, indicating random mixing of large and small particles. After the cycle, however, the distribution peaked sharply near unity, indicating that particles arranged themselves with similarly sized neighbors. This concentration indicates size segregation induced by the compression-decompression cycle.

This size segregation is evident in Fig.~\ref{segregation}(b). Before the cycle, small particles occupied the voids among the large particles, and after the cycle, many of these voids were no longer filled by small particles. This change is closely related to the role of small particles discussed for polydisperse systems, in which the packing density and stiffness depend on whether small particles fill the voids and participate in the contact network~\cite{prasad2017subjamming,petit2020additional,kumar2016tuning}. Regarding the spatial distribution of the particles, the large particles were mainly found in the left half of the figure. In contrast, the small particles were concentrated around the center and the top-right parts, exhibiting a segregation. The position itself had no physical meaning since the periodic boundary condition was imposed. Owing to segregation, the radii of neighboring particles became more similar.

To determine when size segregation develops during the cycle, we calculated the averaged radius ratio between neighboring particles for each configuration,
\begin{equation}
\langle\rho\rangle=\frac{1}{N_{\rm p}}\sum_{(i,j)\in E}\frac{\min(r_i, r_j)}{\max(r_i,r_j)},
\end{equation}
where $E$ is the set of Laguerre-Voronoi neighboring pairs and $N_{\rm p}=|E|$.
As shown in Fig.~\ref{segregation}(d), $\langle\rho\rangle$ first decreased slightly, reaching a minimum at $\phi\simeq1.0$, and then increased from $0.42$ to $0.88$ during compression, which demonstrates that particles acquired neighbors of more similar size.
It indicates a segregation from a homogeneous mixture to a inhomogeneous, segregated state.
The enlarged interval in Fig.~\ref{segregation}(e) shows that some upward changes in $\langle\rho\rangle$ coincide with stress drops. During decompression, $\langle\rho\rangle$ decreased from $0.88$ to $0.80$ and remained far above its initial value, which indicates that the segregation by the cycle with large amplitude persisted during the decompression.
Notably, the increase in $\langle\rho\rangle$ began at $\phi\simeq1.0$, which is close to the range $\phi_{\rm max}\simeq1.0$-$1.2$ in which $\Delta\phi$ reached its minimum and turned to increasing (Fig.~\ref{inflation}(b)).

\subsection{Pressure history}

The reduction in $\phi_{\rm J}$ can be attributed to changes in the local structure induced by size segregation. Numerous studies on the jamming of polydisperse particles~\cite{estrada2016effects,meer2024estimating,shimamoto2023common,voivret2007space,saitoh2025jamming}, including power-law~\cite{estrada2016effects,shimamoto2023common,meer2024estimating,saitoh2025jamming}, exponential~\cite{meer2024estimating}, log-normal~\cite{meer2024estimating}, and $\beta$ distributions~\cite{voivret2007space}, have shown that disordered packings with low polydispersity typically jam near $\phi_{\rm J}\simeq0.84$~\cite{o2002random}, whereas, with high polydispersity, jam at higher densities because small particles fill the voids. Size segregation produces locally narrower size distributions, thereby reducing the local jamming density.
Consequently, segregation homogenizes local structure and reduces $\phi_{\rm J}$. This is consistent with segregation-induced expansion specific to polydisperse systems.

\begin{figure}
    \centering
    \includegraphics[width=8.6cm, bb = 0 0 464 595]{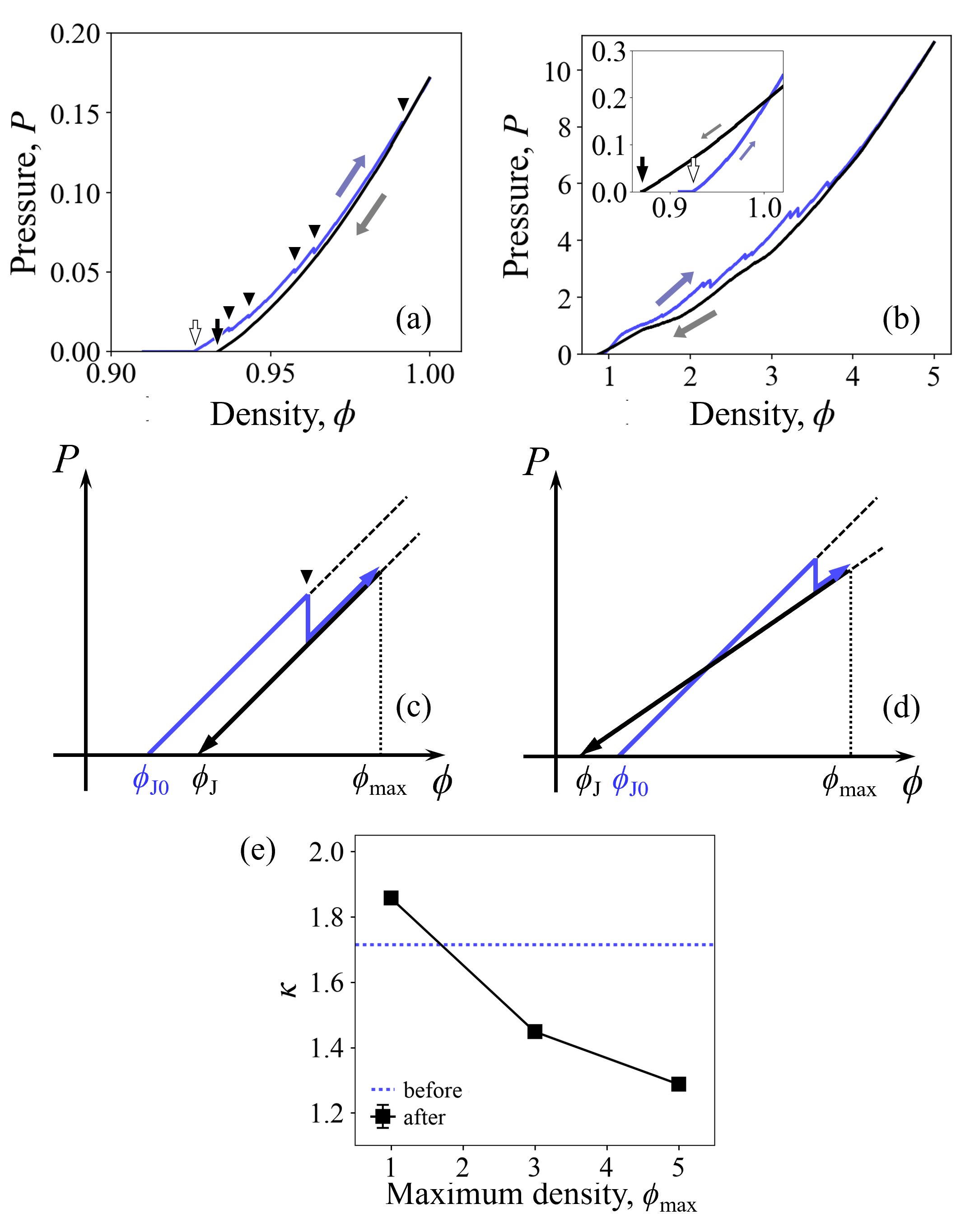}
    \caption{Pressure histories for a power law system with $a=2.306$. (a) Compression (blue curve) and decompression (black curve) for one trial with $\phi_{\rm max}=1$. The initial and final jamming points, $\phi_{\rm J0}$ and $\phi_{\rm J}$, are indicated by white and black arrows, respectively, and the solid triangles indicate pressure drops during compression. (b) Pressure history for a trial with $\phi_{\rm max}=5$. The inset magnifies the low-pressure region and shows that the decompression slope is smaller than the compression slope. (c) Schematic of the increase in $\phi_{\rm J}$ owing to small-amplitude compression. (d) Schematic of large-amplitude compression.
    The effect of the decrease in volumetric modulus exceeds that of stress drops, resulting in a net decrease in $\phi_{\rm J}$. (e) Final volumetric modulus $\kappa$, measured during decompression near the final jamming point, as a function of $\phi_{\rm max}$. The dotted line indicates the modulus measured before the cycle.}
    \label{phi-P}
\end{figure}
The pressure history provides another depiction of both compaction and negative compaction. Figures~\ref{phi-P}(a) and (b) show the pressure history during compression and decompression for $\phi_{{\rm max}}=1$ and $5$, respectively. During compression, the pressure $P$ increased gradually in most steps, whereas it decreased during decompression. Intermittent pressure drops (solid triangles in Fig.~\ref{phi-P}(a)) were also observed during compression. These stress drops correspond to transitions to lower-energy states at the same density caused by plastic particle rearrangements~\cite{maloney2006amorphous,kumar2016memory}. The size segregation shown in Fig.~\ref{segregation} was produced by these configuration changes.
In the high-pressure region ($\phi>4$), the stress drops are rare and $\langle\rho\rangle$ approached a plateau, which indicates that these configuration changes were largely completed, and that few further plastic events occurred. This corresponds to the saturation of $\Delta\phi$ observed in Fig.~\ref{inflation}(a).

Notably, the large-amplitude cycle reduced the pressure slope, as shown in the inset of Fig.~\ref{phi-P}(b). This decrease implies a reduction in the volumetric modulus, $\kappa=dP/d\phi$. The difference in the volumetric modulus was most pronounced in the low-pressure region ($\phi<1$), where the system experienced few plastic events during compression. It was less noticeable in the high-pressure region ($\phi>4$), where the rearrangements had largely been completed. The reduction in $\kappa$ indicates that particle rearrangements and the resulting size segregation significantly alter the system's response to compression. The pressure history therefore suggests that rearrangements caused by large-amplitude compression decrease $\kappa$, contributing to the observed expansion.

The observed stress drops and changes in the volumetric modulus in the pressure history provide a unified framework for understanding both compaction and negative compaction. Compaction can be explained solely by the stress drops. A stress drop signifies a transition to a lower-pressure state with the same density. Thus, when decompression follows such stress drops, the pressure decreases to zero with less decompression, compared to the compression process (Figure~\ref{phi-P}(c)). Consequently, there is an increase in the jamming point, $\phi_{\rm J}$. This established mechanism~\cite{luding2021does} explains the compaction observed in the bidisperse systems or polydisperse systems at small $\phi_{\rm max}$. Our additional finding is that, for the highly polydisperse systems at large $\phi_{\rm max}$, segregation and the associated reduction in $\kappa$ can overcome this compaction mechanism.
\begin{figure}
    \centering
    \includegraphics[width=8.6cm, bb = 0 0 570 400]{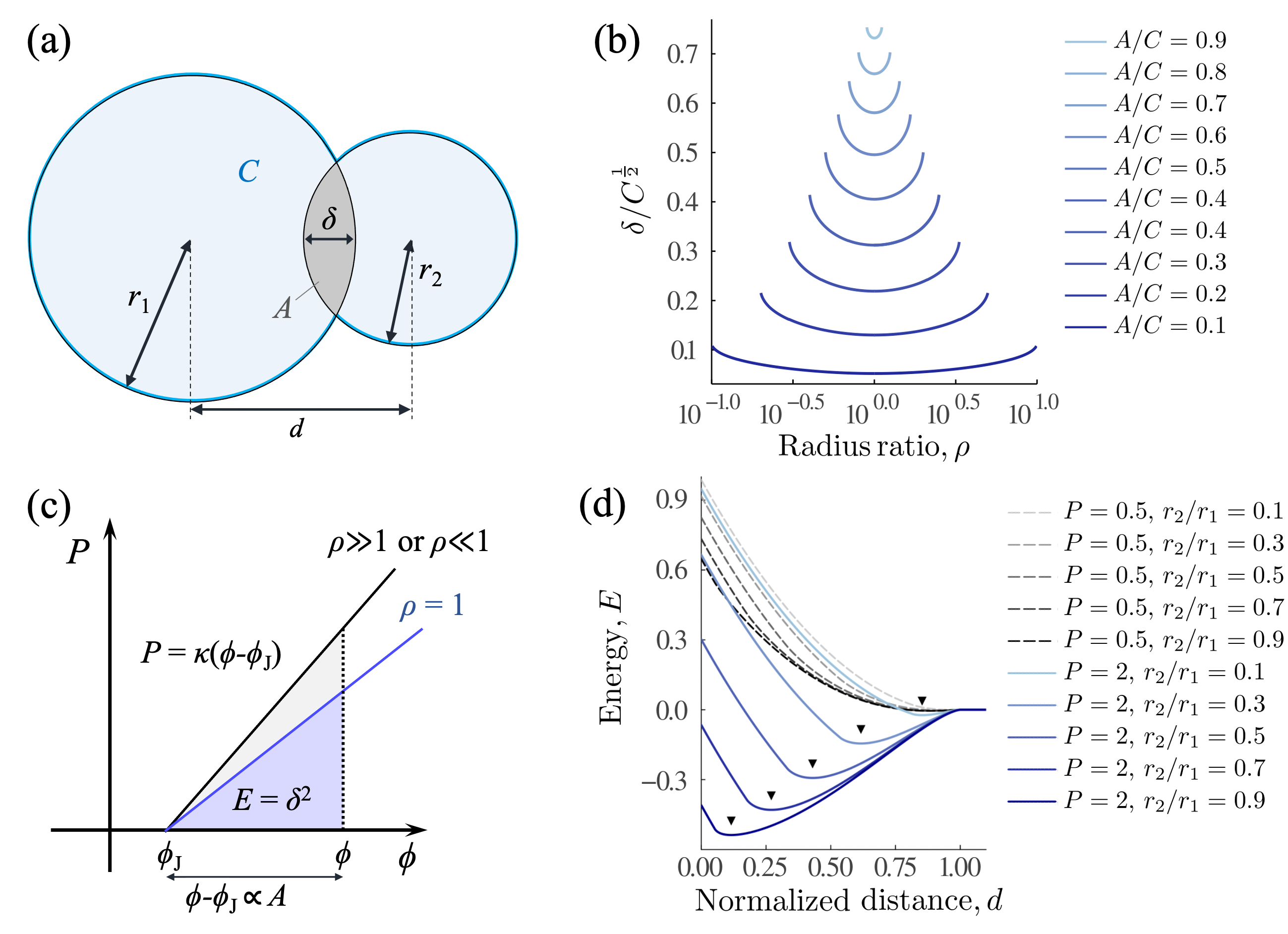}
    \caption{(a) Schematic of particles in contact, showing the excluded volume reduced by the overlap of them in grey.
    (b) The intrusion distance, $\delta=r_1+r_2-d$, of two contacting particles as a function of the ratio of the radii of the two circles, $\rho=r_1/r_2$, for fixed overlapping area, $A$. The length scale is normalized by $C^{\frac{1}{2}}$.
    (c) Schematic of the change in volumetric modulus, $\kappa$, caused by size segregation.
    (d) Effective pairwise potentials for several parameters. Triangles indicate minima. The distance is normalized by $r_1+r_2$.}
    \label{rR}
\end{figure}

In contrast, stress drops alone cannot explain the negative compaction observed in the highly polydisperse systems at large $\phi_{\rm max}$. Large-amplitude compression also considerably reduces the volumetric modulus $\kappa$. Figure~\ref{phi-P}(e) shows $\kappa$ after the cycle as a function of $\phi_{\max}$. It decreased with increasing $\phi_{\max}$ and fell below its initial value before the cycle.
A smaller $\kappa$ requires greater decompression before the pressure reaches zero and therefore reduces $\phi_{\rm J}$ (Fig.~\ref{phi-P}(d)). Thus, while stress drops increase $\phi_{\rm J}$, the simultaneous reduction in $\kappa$ counteracts this effect. Negative compaction occurs when the decrease in $\phi_{\rm J}$ caused by the considerably reduced $\kappa$ outweighs the increase caused by stress drops. 

\section{Discussion}

The present results show that transitions between metastable states can cause expansion in highly polydisperse soft-particle assemblies. This behavior is reminiscent of negative compressibility in metamaterials, in which the volume responds to the stress in the opposite direction to intuition~\cite{nicolaou2012mechanical,ducarme2025exotic,nicolaou2024metamaterials,qu2017experiments}. Whereas the transition in such metamaterials changes the structure inverse against the applied load~\cite{nicolaou2012mechanical,ducarme2025exotic}, the stress drop in our system shifts the system toward compaction. The expansion emerges from the size segregation and reduction in $\kappa.$

\subsection{Volumetric modulus}

The reduction in the volumetric modulus can be explained by a simple model based on changes in the radius ratio between neighboring particles. In this model, two particles with radii $r_1$ and $r_2$ are in contact (Fig.~\ref{rR}(a)). The total particle area $C = \pi(r_1^2 + r_2^2)$ and the overlap area $A$ are fixed, and we calculate the penetration distance $\delta = r_1+r_2-d$ as a function of the radius ratio $\rho = r_1/r_2$. The numerically derived values of $\delta(\rho)$ for various overlap areas $A$ are shown in Fig.~\ref{rR}(b), where the length scale is normalized by $C^{\frac{1}{2}}$ (see Appendix~\ref{App} for analytical solutions). In all cases, $\delta(\rho)$ is convex and reaches its minimum when $r_1 = r_2$. This result suggests that the size segregation, which brings the radii of neighboring particles $\rho$ closer together, leads to a reduction in the volumetric modulus (Fig.~\ref{rR}(c)). In this model, $A$ and $\delta^2$ are analogous to the ``distance'' from the jamming point, $\phi - \phi_{\rm c}$, and the potential energy, $E$, respectively. In the linear regime, where $P = \kappa(\phi - \phi_{\rm c})$, the energy is given by $E = \int_{\phi_{\rm c}}^{\phi} P d\phi' = \kappa/2 (\phi - \phi_{\rm c})^2$, so $\kappa$ is proportional to $E$ for a fixed $\phi - \phi_{\rm c}$. Therefore, the fact that $\delta(\rho)$ is minimized when $\rho = 1$ implies that size segregation reduces $\kappa$ by lowering the system's potential energy.

\subsection{Effective attractions}

The observed size segregation suggests that particles of similar size tend to accumulate during the compression-decompression cycle. This is puzzling, as the underlying inter-particle interactions in our model are purely repulsive and isotropic. To rationalize this counterintuitive behavior, we adopt the effective-interaction perspective based on potential-energy considerations.

In a system near mechanical equilibrium, the configuration of particles minimizes the total potential energy under external constraints. When a particle overlaps with a neighboring one, it locally reduces the excluded volume experienced by surrounding particles. Under pressure, this leads to a reduction in the total potential energy. Thus, although the pairwise force is repulsive, overlapping can effectively lower the system's energy.

To capture this behavior, we construct an effective potential that reflects not only the repulsive pairwise interaction but also the pressure-induced energy gain due to the reduction of the excluded volume. 

The dependence of this effective attraction potential on particle size ratio and pressure is discussed. The effective potential energy $v$ between two particles of radii $r_1$ and $r_2$ ($r_2<r_1$) is given by:
\begin{equation}
    v=
    \left\{
    \begin{alignedat}{4}
        &\left(r_1+r_2-d\right)^2-\pi Pr_2^2 & r_1 > d+r_2,\\
        &0&d > r_1+r_2,\\
        &\left(r_1+r_2-d\right)^2-P&\\
        &\left[{r_1}^2\left(\theta-\frac{\sin{2\theta}}{2}\right)\right.\left.+{r_2}^2\left(\phi-\frac{\sin{2\phi}}{2}\right)\right]& {\rm otherwise,}
    \end{alignedat}
    \right.
    \label{effec_pot}
\end{equation}
where $P$ is the pressure, $d$ is the distance between the particles, and 
the angles $\theta$ and $\phi$ are defined as:
\begin{equation}
    \theta=\arccos{\frac{d^2+r_1^2-r_2^2}{2dr_1}},\;
    \phi=\arccos{\frac{d^2-r_1^2+r_2^2}{2dr_2}}.
\end{equation}
The first term on the right-hand side of Eq.~\ref{effec_pot} represents the repulsive potential between the particles. The second term considers the product of the surrounding particles' pressure and the excluded volume reduced by overlap (illustrated by the gray region in Fig.~\ref{rR}(a)). When the pressure is approximated as hydrostatic, a small inter-particle distance reduces the potential energy due to the surrounding particles' pressure. Consequently, although the particles interact through a purely repulsive force, an effective attraction emerges from this potential energy.

Next, we discuss the configurations of particles under this effective attraction. 
Figure~\ref{rR}(d) presents plots of $v$  described in Eq.~\ref{effec_pot} for various parameter sets, showing the evolution of the interactions under different pressures and particle size ratios. The distance between particles at a mechanical equilibrium corresponds to the value at the minimum of the effective potential $v$.
For a small penetration distance, $\delta$, the overlap area scales as $\delta^{\frac{3}{2}}$, and the potential energy scales as $\delta^2$. The effective potential therefore has the asymptotic form
    \begin{equation}
    v(\delta)\simeq {\epsilon}\delta^2
    -C P\delta^{\frac{3}{2}},
    \label{asympt}
    \end{equation}
where $C$ is a positive factor. Consequently, a shallow minimum exists for any finite pressure.
The position and depth of this minimum scale as
$\delta_{\rm min}\propto P^2$ and $v_{\rm min}\propto-P^4$, respectively. Thus, the effective
attraction becomes continuously stronger as the maximum pressure increases.
At low pressures, the conventional compaction mechanism dominates, whereas at higher pressures, size segregation can overcome this mechanism. We therefore interpret the crossover from compaction to expansion as a competition between these two mechanisms.
At low pressures, the interactions remain almost repulsive (dashed lines in Fig.~\ref{rR}(d)).

However, as the pressure increases, the interactions become effectively attractive, with the potential minima deepening and shifting closer to the origin as the particle size ratio, $r_2/r_1$, approaches unity (solid lines in Fig.~\ref{rR}(d)).
Consequently, although all particles in the system interact through repulsion, particles of similar size effectively attract and accumulate, leading to size segregation in polydisperse mixtures. The argument regarding effective potential provides a microscopic mechanism for the size segregation induced by compression.

We position this effective attractive interaction within the context of known particle interactions. The attraction arises in a damped, gravity-free system, distinguishing it from gravity-driven phenomena such as size segregation in rotating drum experiments~\cite{oyama1939motion} or the Brazil nut effect~\cite{smith1929segregation,breu2003reversing}. However, an analogy can be drawn with the depletion interaction, an attractive force typically observed in thermal systems at the microscopic scale. While the depletion interaction is primarily entropic~\cite{asakura1954interaction,asakura1958interaction}, there are cases where an energetic component plays a role, particularly in systems involving thermal polymers with soft-core repulsion~\cite{benton2012unexpected,sukenik2013balance,sapir2014origin}. The energetic component observed in these systems can be considered an example of the attractive interaction discussed in this study. The attraction described here can be viewed as an energetic counterpart of the depletion interaction. This interpretation situates the attraction within the broader framework of particle interactions, offering a fresh perspective on energy-driven segregation.

Although the present simulations are in two dimensions, the argument on effective potential may be able to be extended to three dimensions, with a qualitatively different results. For two spheres, the overlap volume is $V\simeq\pi\delta^2\frac{r_1r_2}{r_1+r_2}$ for small $\delta$, which is of the same order in $\delta$ as the harmonic repulsion. The three-dimensional counterpart of Eq.~\ref{asympt} is therefore
\begin{equation}
v(\delta)\simeq\left(\epsilon-\pi p\frac{r_1r_2}{r_1+r_2}\right)\delta^2,
\label{v3d}
\end{equation}
where $p$ is the pressure. Whereas a minimum exists at any finite pressure in two dimensions, effective potential develops a minimum only above a threshold pressure, $p_{\rm c}=\epsilon(r_1+r_2)/\pi r_1r_2$ when 3D.
This suggests a qualitative difference in the emergence of effective attraction between two and three dimensions. Whether such an attraction leads to size segregation and expansion in three dimensions remains to be investigated.

The composition of the particle-size distribution provides another control parameter. The bidisperse systems used as reference here are equimolar, and we have not tested the different number or radius ratio of the mixture. Other distributions, such as equal volume, may lead to different results. The onset and magnitude of negative compaction may depend on the distribution.

\section{Conclusion}
In summary, we have shown that a quasistatic cycle of compression and decompression can cause both compaction and expansion in the highly polydisperse disks. By comparing the jamming densities before and after the cycle, we found that small-amplitude cycles increase the jamming density, whereas large-amplitude cycles decrease it. This non-monotonic transition is associated with size segregation, which increases contacts between similarly sized particles and reduces the volumetric modulus of the entire system. We propose this as an additional mechanism operating under broad polydispersity and large-amplitude isotropic cycling, not as the only possible mechanism for dilation or expansion. These results suggest that loading-induced particle rearrangements modify not only particle positions but also the spatial arrangement of particle sizes, thereby changing the jamming density.

More broadly, the present results demonstrate that expansion can emerge in disordered granular systems through a mechanism that differs from those previously identified in structured metamaterials. Rather than arising from a single instability or snapping transition, the expansion originates from cumulative particle rearrangements that transform a homogeneous mixture into an inhomogeneous, segregated state.
Because soft and penetrable interactions arise in dendrimers, star polymers, and particles interacting through ultrasoft long-range potentials~\cite{advzic2024engineering,mladek2008computer,gotze2004tunable,likos2006soft,kolezakis2025microscopic}, the mechanism identified here may provide a route to compression-induced expansion in disordered materials.

\begin{acknowledgments}
    This research was funded by the Japan Society for the Promotion of Science (JSPS) KAKENHI [grant nos. 23KJ0753 (D.S.) and 23K22459 (M.Y.)], Japan Science and Technology Agency (JST) Program [FOREST, JPMJFR213Y (M.Y.) and CREST, JPMJCR2645 (M.Y.)], and Toyota Konpon Research Institute, Inc. D.S. acknowledges the RIKEN Special Postdoctoral Researcher Program.
\end{acknowledgments}

\appendix
\section{Dependence of the penetration distance $\delta$ on the radius ratio $\rho$}\label{App}
In the main text, we discussed the penetration distance, $\delta=r_1+r_2-d$ of two particles in contact. We studied the dependence of $\delta$ on the ratio of radii $\rho=\frac{r_1}{r_2}$ under the constraint that the sum of particle areas $C=r_1^2+r_2^2$ and the overlap area $A$ are fixed, and numerically confirmed that it is minimized when $\rho=1$.

\begin{figure}
    \centering
    \includegraphics[width=8.6cm, bb = 0 0 330 260]{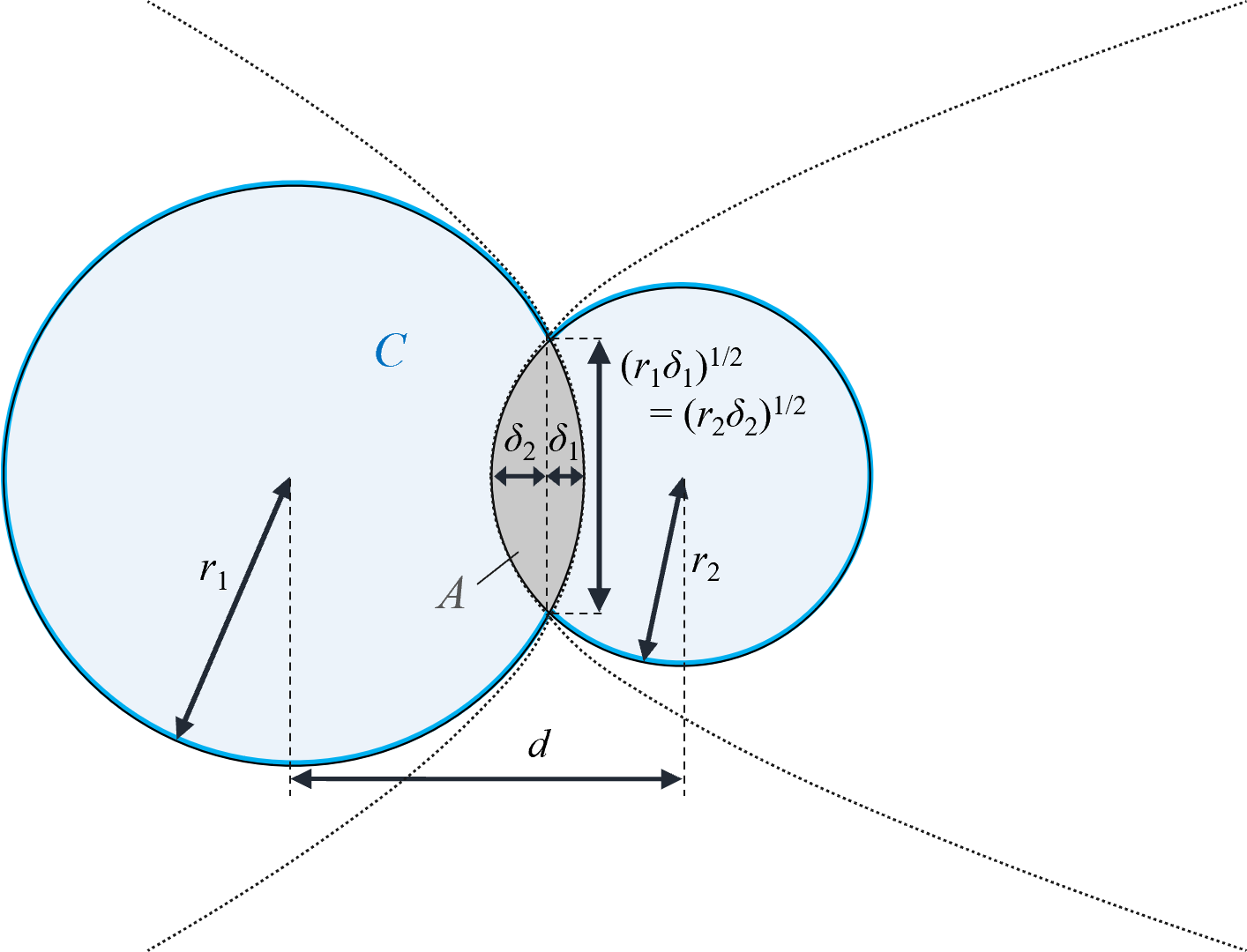}
    \caption{Schematic of overlapping two circles. The dotted curves are parabolas approximating the circumferences.}
    \label{para}
\end{figure}

Here we show that $\rho(\delta)$ takes a minimum value at $\rho=1$ when $\delta$ is sufficiently small by elementary argument. We begin by approximating the surfaces of two particles by parabolas (Fig.~\ref{para}). Under this approximation, the overlapping area of the two particles is expressed as follows:
\begin{eqnarray}
    A&=&\int^{\delta_1}_0 dx \sqrt{r_1x}+\int^{\delta_2}_0 dx \sqrt{r_2x},\\
    &=&\frac{2}{3}\left(r_1^{\frac{1}{2}}\delta_1^{\frac{3}{2}}+r_2^{\frac{1}{2}}\delta_2^{\frac{3}{2}}\right),\\
    &=&\frac{2}{3}\sqrt{r_1\delta_1}\delta.
    \label{A}
\end{eqnarray}
We here used the following relationship, which can be verified to hold from Fig.~\ref{para}:
\begin{eqnarray}
    \delta=\delta_1+\delta_2,\label{delta}
\end{eqnarray}
\begin{eqnarray}
    r_1\delta_1=r_2\delta_2.
    \label{sqrt}
\end{eqnarray}
From the definition of $\rho$ and Eq.~\ref{delta}, we obtain
\begin{eqnarray}
    \delta_1 = \frac{\delta}{\rho+1}.
    \label{delta1}
\end{eqnarray}
From the definition of $C$ and Eq.~\ref{delta} and Eq.~\ref{sqrt}, we obtain
\begin{eqnarray}
    r_1^2=\frac{C}{\frac{1}{\rho^2}+1}.
    \label{r1}
\end{eqnarray}
We eliminate $r_1$ and $\delta_1$ from Eq.~\ref{A} using Eq.~\ref{r1} and Eq.~\ref{delta1}:
\begin{eqnarray}
    \delta^2&=&\left(\frac{3A}{4\sqrt{r_1\delta_1}}\right)^2\\
    &=&\left(\frac{3A}{4}\right)^2\sqrt{\frac{\frac{1}{\rho^2}+1}{C}}\frac{\rho+1}{\delta}
\end{eqnarray}
\begin{eqnarray}
    \delta&=&\left[\frac{9A^2}{16C^{\frac{1}{2}}}\right]^\frac{1}{3}\left[\left(\frac{1}{\rho^2}+1\right)\left(\rho+1\right)^2\right]^\frac{1}{6}\\
    &=&\left[\frac{9A^2}{16C^{\frac{1}{2}}}\right]^\frac{1}{3}(R^2+2R)^\frac{1}{6}
\end{eqnarray}
Here $R=\rho+\rho^{-1}$ and $\rho\geq0$ thus $R\geq2$. This function is convex and takes its minimum value when $R=2$, thus it was shown to be minimum when $\rho=1$.

\bibliography{poly_abrv2}

\end{document}